\documentstyle[12pt]{article} 
\newcommand{\beq}{\begin{equation}}
\newcommand{\eeq}{\end{equation}}
\newcommand{\beqa}{\begin{eqnarray}}
\newcommand{\eeqa}{\end{eqnarray}}
\newcommand{\ba}{\begin{array}}
\newcommand{\ea}{\end{array}}

\begin{document}

\begin{center}
{\Large \bf BEC in Nonextensive Statistical Mechanics: \\
Some Additional Results} 
\vskip 0.8cm 
Luca Salasnich \\
\vskip 0.5cm
Istituto Nazionale per la Fisica della Materia, Unit\`a di Milano,\\ 
Dipartimento di Fisica, Universit\`a di Milano, \\ 
Via Celoria 16, 20133 Milano, Italy 
\end{center}

\vskip 1.5cm

\begin{center}
{\bf Abstract} 
\end{center} 
In a recent paper [Int. J. Mod. Phys. B {\bf 14}, 405 (2000)] 
we discussed the Bose-Einstein condensation (BEC) 
in the framework of Tsallis's nonextensive 
statistical mechanics. In particular, we studied 
an ideal gas of bosons in a confining harmonic potential. 
In this memoir we generalize our 
previous analysis by investigating an ideal Bose gas 
in a generic power-law external potential. 
We derive analytical formulas for 
the energy of the system, the BEC transition temperature 
and the condensed fraction. 

\vskip 0.5cm


\newpage

Recently we analyzed the consequences of weak nonextensivity 
on Bose-Einstein Condensation for an ideal Bose gas$^{1}$ by using 
Tsallis's nonextensive statistical mechanics (NSM).$^{2}$ 
From the nonextensive Bose-Einstein 
distribution$^{3-5}$ we derived the BEC transition temperature, 
the condensed fraction and the energy per particle 
in three different cases: the homogeneous gas, the gas in a harmonic trap and 
the relativistic homogenous gas. 
\par
In this short contribution we generalize our 
previous results by considering the ideal Bose gas 
in a generic power-law potential. All the calculations are 
performed by assuming a D-dimensional space. 
\par 
The NSM predicts that, 
for an ideal quantum gas of identical bosons in the grand canonical ensemble, 
the weak nonextensivity formula of the average number 
of particles with energy $\epsilon$ is given by 
\beq
\langle n(\epsilon )\rangle_q = 
{1\over e^{\beta(\epsilon -\mu)}-1}  
+ {1\over 2} (q-1) {\beta^2 (\epsilon - \mu)^2 e^{\beta(\epsilon -\mu)} 
\over (e^{\beta(\epsilon -\mu)}-1)^2 } \; , 
\eeq
where $\mu$ is the chemical potential and $\beta=1/(kT)$ 
with $k$ the Boltzmann constant and $T$ the temperature.$^{3}$ 
The parameter $q$ is a measure of the lack of extensivity: 
for $q=1$ one recovers the familiar Bose-Einstein distribution. 
An interesting interpretation of $q$, in terms of fluctuations 
of the parameters which appear in the standard exponential distribution, 
can be found in Ref. 6. 
\par
The total number $N$ of particle and the energy $E$ 
for a system of non-interacting bosons can be written 
\beq
N=\int_0^{\infty} 
d\epsilon \; \rho(\epsilon ) \; \langle n(\epsilon )\rangle_q \; 
\hskip 0.7cm \hbox{and} \hskip 0.7cm 
E= \int_0^{\infty} d\epsilon \; 
\epsilon \; \rho(\epsilon ) \; \langle n(\epsilon )\rangle_q \; ,  
\eeq 
where $\rho (\epsilon )$ is the density of states. 
It can be obtained from the formula 
\beq
\rho(\epsilon ) = \int {d^D{\bf r} d^D{\bf p}\over (2\pi\hbar)^D} 
\delta (\epsilon - H({\bf p},{\bf r})) \; ,  
\eeq
where $H({\bf p},{\bf r})$ is the classical single-particle 
Hamiltonian of the system in a D-dimensional space. 
If $H({\bf p},{\bf r})={\bf p}^2/(2m)+U({\bf r})$ then 
it is easy to show that 
\beq 
\rho(\epsilon ) = 
\left({m\over 2\pi \hbar^2}\right)^{D\over 2} 
{1\over \Gamma({D\over 2})} 
\int d^D{\bf r} \; 
\left( \epsilon - U({\bf r}) \right)^{(D-2)\over 2} 
\; ,  
\eeq
where $\Gamma(x)$ is the factorial function. 
\par
Let us consider the power-law potential given by 
\beq
U({\bf r})= A \; r^{\alpha} \; ,
\eeq 
where $r=|{\bf r}|$ and $\alpha$ is the exponent. 
The study of this potential is useful to analyze the 
effects of adiabatic changes in the confining trap.$^{7}$ 
The density of states can be calculated from the two previous formulas 
and reads 
\beq 
\rho(\epsilon ) = 
\left({m\over 2\hbar^2}\right)^{D\over 2} 
\left({1\over A}\right)^{D\over \alpha} 
{\Gamma({D\over \alpha}+1) 
\over \Gamma({D\over 2}+1) 
\Gamma({D\over 2} +{D\over \alpha})} \epsilon^{{D\over 2} 
+{D\over \alpha} -1}  
\; . 
\eeq 
\par
At the BEC transition temperature, 
the chemical potential $\mu$ is zero and 
at $\mu=0$ the number $N$ of particles 
and energy $E$ 
can be analytically determined. 
The number of particles is given by 
$$
N = \left( kT \right)^{{D\over 2} + {D\over \alpha}} 
\left({m\over 2 \hbar^2}\right)^{D\over 2} 
\left({1\over A}\right)^{D\over \alpha} 
{\Gamma({D\over \alpha}+1) 
\zeta({D\over 2} + {D\over \alpha}) 
\over \Gamma({D\over 2}+1) } \times 
$$
\beq
\times \left[1 + {1\over 2}(q-1) 
{\Gamma({D\over 2} + {D\over \alpha}+2) \zeta({D\over 2} 
+ {D\over \alpha}+1)\over 
\Gamma({D\over 2} + {D\over \alpha}) \zeta({D\over 2} 
+ {D\over \alpha}) } \right] \; 
\eeq 
and the energy satisfies the following relation 
$$
{E\over k T} = \left(kT\right)^{{D\over 2} + {D\over \alpha}} 
\left({m\over 2\hbar^2}\right)^{D\over 2} 
\left({1\over A}\right)^{D\over \alpha} 
{({D\over 2} + {D\over \alpha}) 
\Gamma({D\over \alpha}+1) \zeta({D\over 2}+ {D\over \alpha}+1)
\over \Gamma({D\over 2}+1) } 
\times 
$$
\beq
\times \left[1 + {1\over 2}(q-1) 
{\Gamma({D\over 2} + {D\over \alpha}+3) \zeta({D\over 2} 
+ {D\over \alpha} +2)\over 
\Gamma({D\over 2} + {D\over \alpha}+1) \zeta({D\over 2} 
+ {D\over \alpha} +1) } \right] \; ,
\eeq 
where $\zeta(x)$ is the Riemann $\zeta$-function. 
Note that our formula of the energy can be 
easily generalized above the critical temperature by substituting 
the Riemann function $\zeta(D)$ with the polylogarithm 
function $Li_{D}(z)=\sum_{k=1}^{\infty}z^k/k^D$, that depends 
on the fugacity $z=e^{\beta \mu}$. 
From the energy one can easily obtain the specific heat and the other 
thermodynamical quantities. 
\par
By inverting the function $N=N(T)$ one finds 
the transition temperature $T_q$. It is given by 
$$
kT_q = \left[ 
\left({2 \hbar^2\over m}\right)^{D\over 2} 
A^{D\over \alpha} 
{\Gamma({D\over 2}+1) \over \Gamma({D\over \alpha}+1) 
\zeta({D\over 2} + {D\over \alpha})} N 
\right]^{1\over {D\over 2}+{D\over \alpha}} \times 
$$
\beq
\times \left[1 + {1\over 2}(q-1) 
{\Gamma({D\over 2} + {D\over \alpha}+2) \zeta({D\over 2} 
+ {D\over \alpha} +1)\over 
\Gamma({D\over 2} + {D\over \alpha}) 
\zeta({D\over 2} + {D\over \alpha}) } 
\right]^{{D\over 2}+ {D\over \alpha}}  
\; . 
\eeq 
The Eqs. (6)--(9) generalize the results obtained 
in our previous paper.$^{1}$ 
In fact, by setting $\alpha =2$ and $A=m\omega^2r^2/2$ one recovers 
the formulas for the Bose gas in a harmonic trap. 
The results for a rigid box are instead obtained by letting 
${D\over \alpha}\to 0$, where the density of particles 
per unit length is given by $N/\Omega_D$ and 
$\Omega_D= D\pi^{D\over 2}/
\Gamma({D\over 2}+1)$ is the volume of the D-dimensional 
sphere. Obviously, for $q=1$ one gets the equations 
of the extensive thermodynamics.$^{8}$ 
In particular, with $q=1$ and $D=3$, the Eq. (9) 
gives the extensive BEC transition temperature obtained by Bagnato, 
Pritchard and Kleppner.$^{9}$ 
\par 
The Eq. (9) shows that the critical temperature 
$T_q$ grows by increasing the nonextensive parameter $q$. 
Moreover, one observes that, because $\zeta(1)=\infty$, 
BEC is possible if and only if the following condition is 
satisfied  
\beq 
{D\over 2}>1-{D\over \alpha} \; .
\eeq 
Thus, we have found a remarkable relation between the space dimension $D$ 
and the exponent $\alpha$ of the confining power-law potential. 
For example, for $D=2$ there is no BEC in a homogenous gas 
(${D\over \alpha} \to 0$) but it is possible in a harmonic trap 
($\alpha =2$). Instead, for $D=1$ BEC is possible with $1 < \alpha < 2$ 
(see also Ref. 8). 
\par 
Below $T_q$, a macroscopic number $N_0$ of particle occupies 
the single-particle ground-state of the system. It follows 
that Eq. (7) gives the number $N-N_0$ of non-condensed particles and 
the condensed fraction reads 
\beq
{N_0\over N}=1-\left({T\over T_q}\right)^{{D\over 2}+{D\over \alpha}} \; . 
\eeq 
\par 
In conclusion, we have analyzed the consequences of 
Tsallis's nonextensive statistical mechanics for an ideal 
Bose gas confined in a power-law potential. 
We have obtained analytical formulas for 
the energy of the system, the BEC transition temperature 
and the condensed fraction. In the appropriate limits, 
such formulas reduce to results found in previous papers. 
Moreover, we have shown that BEC is possible if and only if 
${D\over 2}>1-{D\over \alpha}$, where $D$ is 
the space dimension and $\alpha$ is 
the exponent of the confining power-law potential. 

\newpage 

\section*{References}

\begin{description}

\item{\ 1.} L. Salasnich, Int. J. Mod. Phys. B {\bf 14}, 405 (2000). 

\item{\ 2.} C. Tsallis, J. Stat. Phys. {\bf 52}, 479 (1988). 

\item{\ 3.} E.M.F. Curado and C. Tsallis, J. Phys. A {\bf 24}, L69 (1991) 

\item{\ 4.} F. Buyukkilic, D. Demirhan, and A. Gulec, Phys. Lett. A 
{\bf 197}, 209 (1995) 

\item{\ 5.} Q.A. Wang and A. Le Mehaute, Phys. Lett. A {\bf 235}, 
222 (1997). 

\item{\ 6.} G. Wilk and Z. Wlodarczyk, Phys. Rev. Lett. {\bf 84}, 2770 (2000). 

\item{\ 7.} F. Dalfovo, S. Giorgini, L.P. Pitaevskii, and S. 
Stringari, Rev. Mod. Phys. {\bf 71}, 463 (1999). 

\item{\ 8.} L. Salasnich, J. Math. Phys. {\bf 41}, 8016 (2000). 

\item{\ 9.} V. Bagnato, D. Pritchard and D. Kleppner, 
Phys. Rev. A {\bf 35}, 4354 (1987). 

\end{description}

\end{document}